# Furtive Quantum Sensing Using Matter-Wave Cloaks

Romain Fleury, and Andrea Alù[*]

Dept. of Electrical and Computer Engineering, The University of Texas at Austin

*alu@mail.utexas.edu

*We introduce the concept of furtive quantum sensing, demonstrating the possibility of concealing quantum objects from matter-waves, while maintaining their ability to interact and get excited by the impinging particles. This is obtained by cloaking, with tailored homogeneous metamaterial layers, quantum systems having internal degrees of freedom. The effect is a low-observable quantum sensor with drastically reduced elastic scattering and optimized absorption levels, a concept that opens interesting venues in particle detection, high-efficiency electrical pumping and quantum supercomputing.*

PACS: 03.75.-b, 78.67.Pt, 85.35.Be, 34.80.Gs, 73.63.Hs.

The enticing possibility of inducing invisibility with passive metamaterial coatings has drawn considerable attention over the past few years, leading to the formulation of different approaches for electromagnetic [1-13], acoustic [14-24], and more recently matter waves [25-33]. Among these techniques, the transformation-optics method redirects the impinging wave around a concealed region, in such a way that any form of scattering is eliminated [1-5,14-21,25-30]. This strategy takes advantage of the fact that a given coordinate transformation of the wave equation is equivalent to a suitable mapping of the constitutive parameter form and their distribution in space, leading to specific recipes for inhomogeneous and anisotropic cloaks that can completely divert, in the ideal case, an impinging wave around a given volume without perturbing it outside the cloak. As an alternative method, the scattering cancellation approach utilizes homogeneous and isotropic shells to suppress the dominant scattering terms in the scattered field expansion, leading to substantial scattering reduction for moderately sized objects, with arguably simpler practical designs [6-13,22-24,31-33]. Both methods have been recently applied to matter waves [25-33], showing that it may be possible to drastically suppress elastic particle scattering off a quantum object by suitably coating it with metamaterials with specifically tailored effective mass $m$ and potential $V$.

An intrinsic peculiarity of the scattering cancellation technique is that the impinging field is allowed to enter the cloaking layer, enabling controlled interaction with the hidden object that is not isolated by the cloak. By placing an absorptive object inside the cloak, it has been theoretically suggested [34], and recently proven experimentally [35], that it may be possible to extract some portion of the impinging energy inside the cloak while remaining essentially undetectable and minimizing the perturbation of the probed field. In the electromagnetic and acoustic scenarios, this has naturally led to the idea of cloaked sensors, characterized by a dramatic reduction of their scattering cross-section and optimized ratio of absorption over scattering [34-47]. This concept has been later extended also to a modified form of transformation-based cloaking, and it has opened exciting venues in near-field scanning optical microscopy [39,40,42], low-observable antennas [44], improved photodetectors [35] and microphones [47].

In this Letter, we show that this concept may be applied to the quantum world, providing the possibility of controlling the ballistic flow of quantum particles through scattering centers and extending the reach of this technology to cold-atom physics, solid-state devices, nanophotonics and quantum computing. With the development of new approaches to quantum electronics dealing with increasingly smaller length scales, the control and monitoring without interference of the ballistic electronic flow has become more and more challenging, and the development of novel strategies for low-observable particle flow sensing is becoming increasingly important. Motivated by these issues, we introduce the concept of furtive quantum sensing, showing that, by surrounding a small quantum system with homogeneous cloak layers with tailored material properties, it may be possible to cancel the elastic scattering while preserving the probability of inelastic scattering events, opening the possibility of performing sensing operations on particle beams without sensibly disturbing them.

Consider the geometry of Figure 1(a), in which a quantum system with internal degrees of freedom, such as a two-level quantum dot or a molecule, is used to detect the presence of impinging particles with defined momentum, such as electrons or cold atoms. The impinging particles can excite the quantum sensor, scattering through one of the available inelastic channels. When this happens, we are able to detect the impinging wave by, for instance, measuring the spontaneous light emission from the quantum system. The interaction, however, also takes the form of elastic scattering, perturbing the momentum direction. Seen from the perspective of the impinging beam, the inelastic interaction represents a loss of a portion of the impinging particles, which are converted to other forms of energy, associated with the absorption cross-section $\sigma_{abs}$ of the system, whereas the elastic scattering contributes to its elastic cross-section $\sigma_{el}$. Quantum sensors are generally chosen such that not all impinging particles participate to inelastic processes, in order to be able to further interact with the matter-wave flow. Therefore, the bare sensor is usually designed to have a small absorption cross-section, which leads to small absorption efficiencies, defined as the ratio between absorption and scattering cross-sections. It would be ideal to be able to suppress the elastic portion of the scattered beam and increase the overall efficiency by avoiding unnecessary contributions to the elastic scattering that divert a significant portion of the impinging beam in all other directions. We envision to achieve this 'ideal quantum sensor' in Figure 1.(b), by surrounding our sensor with a quantum metamaterial cloak with isotropic properties, designed by applying the scattering cancellation technique [31-33] to suppress the overall elastic scattering at energy $E$ of the impinging particles. Because the wave function enters the cloak, the absorption cross-section of the target, i.e., its sensing ability, may be preserved, but the sensor may become significantly less detectable. The scattering efficiency in this case may be made extremely high, essentially unchanging the level of absorption and making this design considerably less intrusive than the bare sensor.

It may be wondered whether this counterintuitive concept may even be allowed by the laws of physics. In order to formally approach the problem, consider the partial wave formulation of quantum scattering from a central field, in which absorption losses are taken into account by considering complex phase shifts $\delta_l$, or equivalently complex scattering coefficients $S_l = e^{2i\delta_l}$. The scattered amplitude at an angle $\theta$ is [48]

$$f_{el}(\theta) = \frac{i}{2k_0} \sum_{l=0}^{+\infty} (2l+1)(1-S_l)P_l(\cos\theta) \tag{1}$$

and the total elastic scattering cross-section may be written as

$$\sigma_{el} = \frac{\pi}{k_0^2} \sum_{l=0}^{+\infty} (2l+1)|1-S_l|^2,  \qquad (2)$$

whereas the total absorption cross-section, taking into account inelastic and reaction scattering is

$$\sigma_{abs} = \frac{\pi}{k_0^2} \sum_{l=0}^{+\infty} (2l+1)(1-|S_l|^2). \qquad (3)$$

The total, or extinction, cross-section $\sigma_{tot} = \sigma_{el} + \sigma_{abs}$ quantifies the number of particles deviating from the direction $\vec{k}_0$ of the incident beam, linked to the forward elastic scattering amplitude $f_{el}(0)$ by the optical theorem [48]

$$\sigma_{tot} = \frac{4\pi}{k_0} \operatorname{Im} f_{el}(0). \qquad (4)$$

This fundamental relation tells us that absorption is necessarily accompanied by elastic scattering and it is impossible to *totally* suppress the elastic scattering of an absorptive target. This is natural, since any transition in the cloak implies taking out one or more of the impinging particles, which would be missing at the back of the sensor (forward direction). A trade-off is therefore expected between absorption and scattering levels, limiting the total elastic scattering reduction achievable with a cloaked quantum sensor given the desired absorption level. This is also consistent with our related findings on cloaked electromagnetic sensors [36-38]. To quantify the fundamental bound between overall absorption and absorption efficiency here, consider Eq. (3) in the case of passive scatterers ($\sigma_{abs} \geq 0$). As a direct consequence of passivity, each coefficient $S_l$ necessarily belongs inside the unit circle in the complex plane and, by studying the parametric dependence of Eqs. (2)-(3) on the scattering coefficients, it is possible to derive a specific range of achievable values for the pair formed by the partial absorption cross-section $\sigma_{abs,l}$ and the partial scattering efficiency $\sigma_{abs,l}/\sigma_{el,l}$ for any scattering order $l$. As an example, the fundamental bound for the *s*-wave ($l=0$) is given by the solid lines in Figure 2, which separates the allowed from the forbidden (shadowed) region. This boundary is formed by two line segments that meet at the point of maximum (resonant) absorption, at which $\sigma_{abs,l} = \sigma_{el,l}$ [48]. If we can accept an absorption level lower than the ideal maximum, as usual in sensors, we may significantly increase the absorption efficiency by operating along the red line and, for reasonably large absorption values, still capable of triggering transitions in our system, we may achieve furtive quantum sensing with significantly reduced scattering. Similar bounds may be derived for any scattering order using Eqs. (2)-(3), essentially quantifying the fundamental limits on quantum furtive sensing.

Consider now the specific scenario of a beam of quantum particles scattered from an object with size smaller than the De-Broglie wavelength $\lambda_0$, such that the *s*-partial wave dominates the scattering. This may occur for sufficiently cold particles or for sufficiently small quantum sensors. We model the sensor as a simple *s*-resonator, characterized by its Breit-Wigner resonance lineshape with resonance energy $E_R$, elastic scattering linewidth $\Gamma_{el}$ and inelastic scattering linewidth $\Gamma_{abs}$. The presence of the sensor can be rigorously modeled in its surrounding region by imposing a suitable condition on the logarithmic derivative of the radial wave function at an arbitrary spherical boundary of radius $a$ larger than the sensor [49]. Our results in [33] for matter-wave cloaking show that in the long-wavelength limit the cloaking condition for *s*-waves only involves the cloak potential, while its effective mass may be independently tailored to suppress *p*- scattered waves. By controlling the potential using metamaterial layers, therefore, we may be able to significantly reduce the elastic scattering from the *s*-resonator. We assume a spherical cloak with homogeneous potential $V_c$ between the radii $a_{c1}$ and $a_{c2}$, and a core potential $V_{in}$ inside this shell. By solving the scattering equations and imposing proper boundary conditions, it is possible to solve for the scattering coefficients $S_l$ [49] and derive the cross-sections using (2)-(3).

As an example, Figure 2 shows the effect, on the electron scattering, of varying the potential cloak for a quantum sensor with $E_R = 0.15\,\text{meV}$, $\Gamma_{el} = 10^{-5}\,\text{eV}$, $\Gamma_{abs} = 10^{-6}\,\text{eV}$ and $a = 1\,\text{Å}$, and a cloak design with $a_{c2} = 1\,\text{nm}$, $a_{c1} = 0.8\,\text{nm}$, for an exciting beam with energy $E = 0.1\,\text{meV}$. The cloak potential $V_c$ is swept over a wide range of values, for two fixed values $V_{in} = 0$ and $V_{in} = -140\,\text{meV}$, generating the two dashed lines. The green point corresponds to the uncloaked case $V_c = 0$, for which the scattering efficiency is relatively low and definitely not optimal (far from our derived bound), showing a strong probability of elastic scattering events compared to inducing an inelastic transition. The value of absorption is significantly lower than the maximum, because the resonator is off-resonance, allowing for large tuning of the scattering efficiency, according to our derived bounds. Indeed, by using a cloak potential $V_c$ with positive (blue dots in the figure) or negative values (black dots) it is possible to drastically modify the absorption cross-section and the scattering efficiency, reaching any value between the purple and red bounds. In particular, by introducing a positive $V_c = 140\,\text{meV}$, we dramatically boost the absorption efficiency (over five orders of magnitude), moderately reducing the value of absorption, but at the same time drastically suppressing the overall scattering. This represents the optimal design for this specific absorption level, and it results in a low-observable, non-intrusive, furtive sensor at the design energy $E$. Different cloak designs would produce different dispersion in Fig. 2, but for each design an optimal value lying in the red portion of the absorption efficiency bound may be derived, providing large flexibility in choosing the level of desired optimal absorption efficiency and the corresponding level of absorption. As a second example, by adjusting the inside potential $V_{in} = -140\,\text{meV}$, we may be able to achieve strongly enhanced absorption efficiency and even increase the overall absorption level, as for the red point in Figure 2.

Further insights into the cloaking mechanism may be grasped by inspecting the energy dependence of the scattering and absorption cross-sections of the system in the uncloaked and cloaked cases, as shown in Figure 3. As expected, a resonant lineshape is observed for the bare *s*-resonator (Fig. 3a), with a maximum peak for both scattering and absorption at its resonant energy $E_R$. Scattering is consistently larger than the absorption within the considered range. When the sensor is surrounded by the optimized cloak with $V_{in} = 0$ and $V_c = 140\,\text{meV}$ (Fig. 3b), a drastic reduction of elastic cross-section is observed around the working energy, producing an essentially invisible quantum sensor, whose absorption remains comparable, albeit lower, than the bare sensor's one. Panel (c) presents a design with reduced scattering and *larger* absorption than the bare sensor, obtained with $V_{in} = -140\,\text{meV}$ and $V_c = 1034\,\text{meV}$. In both cases we achieve considerable scattering reduction despite the covered object being ten times larger than the bare sensor. At higher energy levels, the cloaked system supports a resonance, with a peak in absorption and scattering, and past that energy the absorption efficiency is drastically lowered. The matter-wave cloak provides large flexibility in tailoring the overall absorption at the energy of interest, allowing at the same time optimal absorption efficiency, consistent with our derived bounds. These findings are analytically similar, although physically very different, than the electromagnetic scenario outlined in [7].

We have confirmed our findings using 3D finite-element simulations based on Comsol Multiphysics, modeling the *s*-resonator with a logarithmic derivative surface boundary condition. Figure 4 shows the vector distribution of the time-averaged probability current distribution and the real part of the quantum wave function in the near-field of the sensor, for plane wave incidence from the left. The bare sensor significantly perturbs the matter wave flow (Fig. 4a), but a properly designed cloak (Fig. 4b) restores the planar wave fronts and the straight particle current lines. This results in a major elastic scattering reduction without significantly affecting its ability to sense, as demonstrated by the probability flow inside the cover. An observer located outside the cloak would not be able to easily detect the presence of the sensor, unless performing an extremely large number of quantum experiments based on particle detection, as the probability distribution is essentially equal to the case of no sensor. The beam travels as if the sensor were not there, except that occasionally one particle goes missing, efficiently captured by the sensor, which is tailored to suppress all the unnecessary scattering in all other directions other than its shadow.

To conclude, we have explored here the concept of furtive quantum sensing, the quantum counterpart of cloaked electromagnetic and acoustic sensors. We have shown that, consistent with these other wave scenarios, it is possible to manipulate quantum scattering and drastically reduce the elastic cross-section of a quantum system with internal degrees of freedom, while maintaining its ability to perform sensing via inelastic scattering events. This is achieved using a homogeneous cloak that can cancel the portion of elastic scattering, usually dominant, that is not strictly necessary to sustain the sensing operation. We have derived the fundamental bounds for this concept in terms of absorption efficiency versus total absorption and have provided specific examples applied to a subwavelength *s*-resonator. Our theory may be readily extended to larger sensors and more complex situations involving multiple partial waves by considering multilayered cloaks with tailored effective masses and potentials. Our work extends the reach of sensor cloaking to quantum mechanics, providing new tools for extreme manipulation of matter waves and paving the way to furtive quantum sensing strategies for ballistic particle control and detection at the nanoscale, of interest in efficient electrical pumping, non-intrusive particle detection and quantum computing. This work has been supported by the AFOSR YIP award No. FA9550-11-1-0009 and the DTRA YIP award No. HDTRA1-12-1-0022.

**Figures**

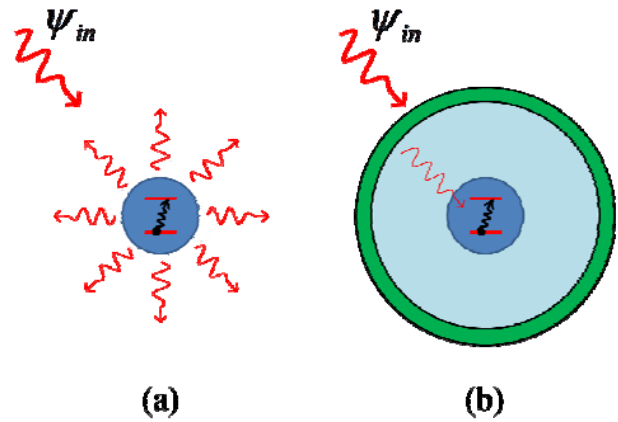

Figure 1 - (a) A bare quantum system with internal degrees of freedom may be able to extract energy and sense an impinging particle beam, but will also perturb the matter wave flow, becoming detectable by its surroundings and affecting its measurement; (b) a cloaked quantum sensor may still be able to extract energy, but with much reduced elastic scattering, making it almost invisible to the particle flow.

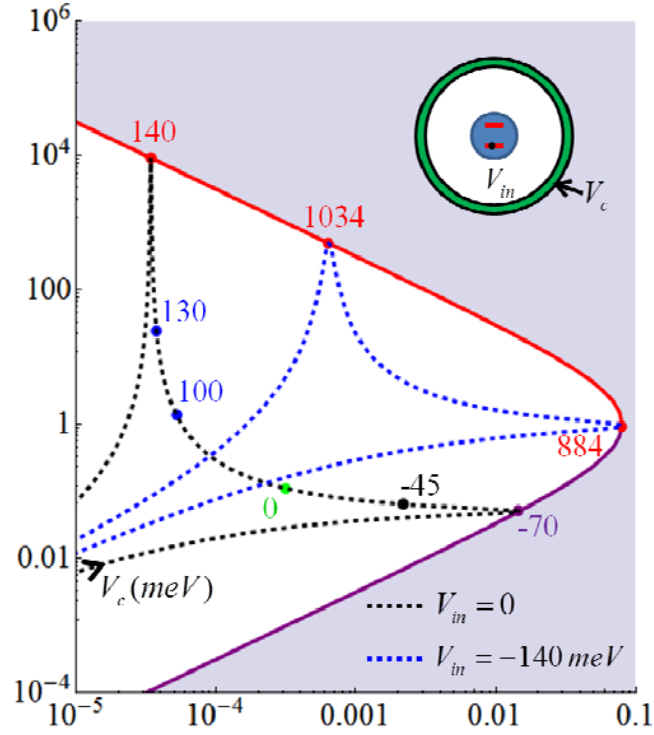

Figure 2– Effect of a core-shell isotropic potential cloak on the electronic scattering and absorption cross-sections of the considered *s*-resonator at $E = 0.1\,\text{meV}$. The uncloaked *s*-resonator has a poor scattering efficiency (green point). By tailoring the cloak properties, it is possible to boost the scattering efficiency by up to five orders of magnitude ($V_{in} = 0$, red point) or reach any optimal absorption level lying on the red line ($V_{in} = -140\,\text{meV}$, red point). The red and purple lines represent fundamental physical bounds for passive *s*-resonators.

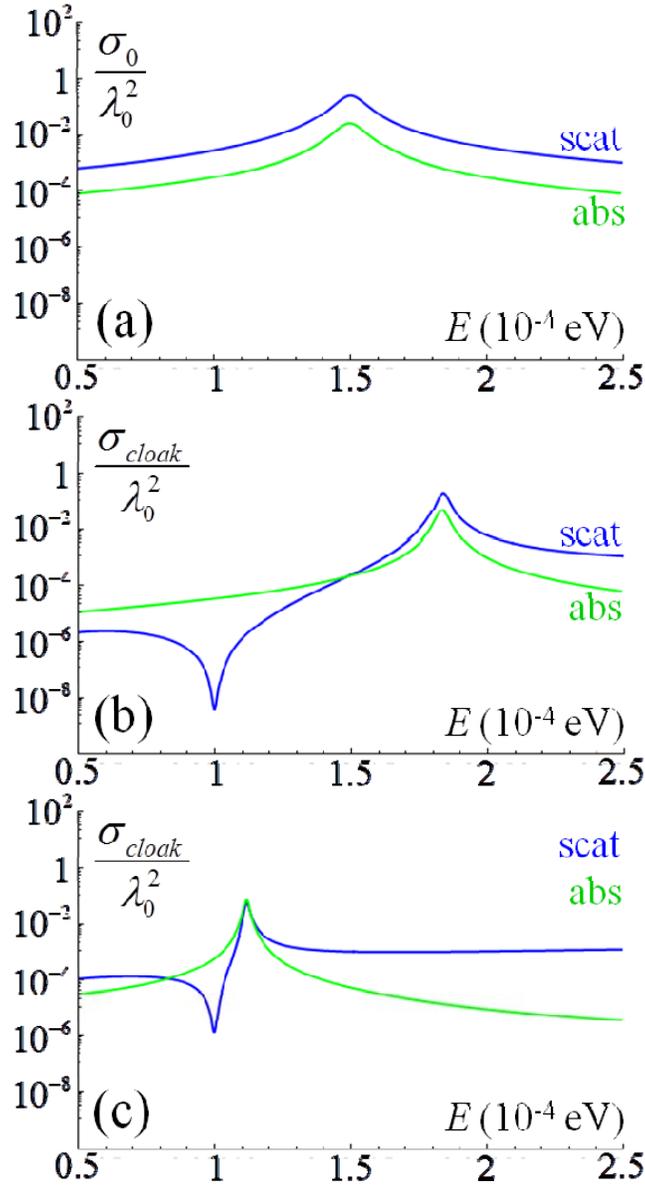

Figure 3 – Scattering and absorption cross-sections versus electron energy for (a) the bare *s*-resonator (green point in Figure 2), (b) a cloaked *s*-resonator with lower absorption than the bare sensor (red point on black dashed line in Figure 2), and (c) a cloaked *s*-resonator with higher absorption than the bare sensor ($V_c = 1034$ meV, blue dashed line in Figure 2).

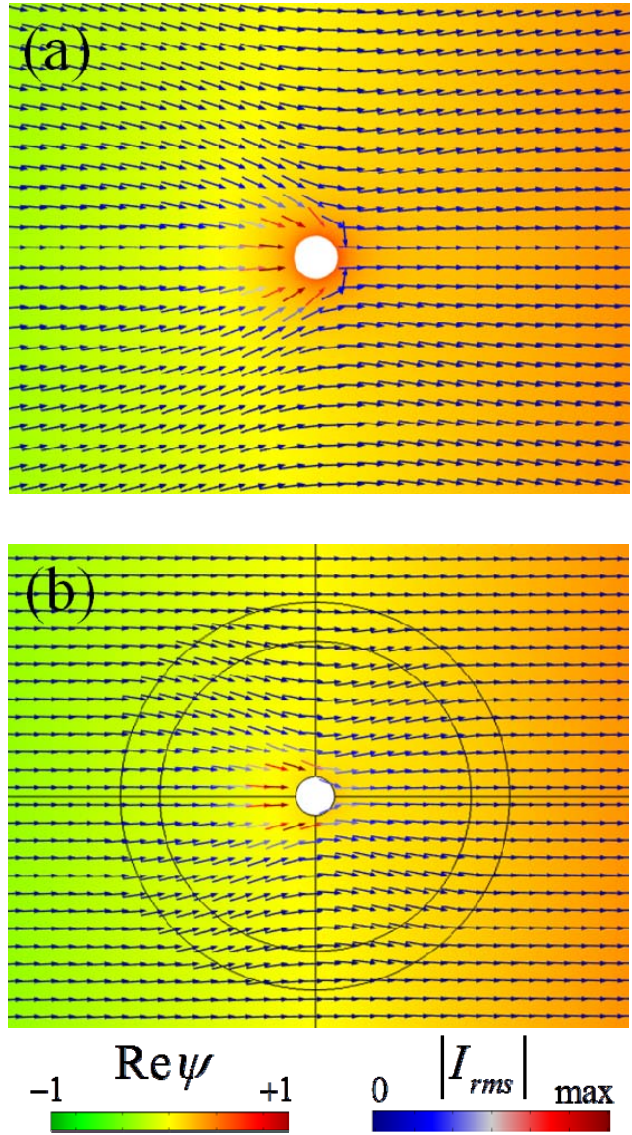

Figure 4 – Finite-element simulation for the electronic quantum scattering for (a) the bare *s*-resonator (green point in Figure 2) and (b) an optimally cloaked *s*-resonator (red point on black dashed line in Figure 2), for $E = 0.1\,\text{meV}$. The color contours show the real part of the wave function and the arrows represent the root-mean-square averaged probability current distribution.

**Online Supplementary Material for the article "Furtive Quantum Sensing Using Matter-Wave Cloaks", by R. Fleury and A. Alù**

In this supplementary material, we analytically solve the quantum scattering problem for a cloaked *s*-resonator, as considered in our work. The geometry is shown is Fig. S-1, in which an impinging quantum particle beam of energy $E$ and defined momentum is scattered off a quantum *s*-resonator, placed in a homogeneous spherical core-shell potential structure with core potential $V_{in}$, and shell potential $V_c$ between the radial distances $r = a_{c1}$ and $r = a_{c2}$. The *s*-resonator is modeled by imposing the logarithmic derivative on a spherical mathematical surface enclosing it, located at $r = a$. For simplicity, we neglect here any spin-dependent effect. The quantum wave function satisfies the non-relativistic time-independent Schrödinger equation

$$\left[\frac{-\hbar^2}{2m_0}\nabla^2 + V(r)\right]\psi = E\psi \qquad (S.1)$$

where $m_0$ is the constant (positive) effective mass of the incident particles, and $V(r)$ is the potential energy, assumed to be zero everywhere except inside the core and the shell, where it takes the values $V_{in}$ and $V_c$.

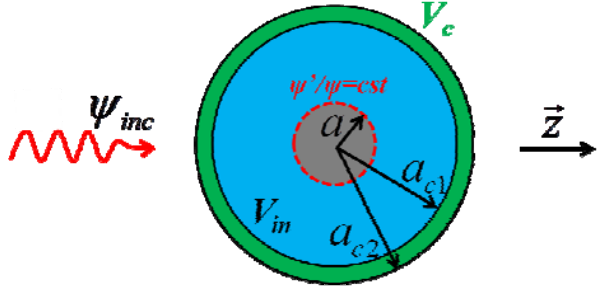

Figure S1- Geometry of the problem under analysis: an *s*-resonator (modeled via a logarithmic derivative boundary condition on the dashed red surface), inside a spherical core-shell potential structure.

The impinging plane wave, assumed to be traveling in the $\vec{z}$ direction, can be expanded in spherical waves in the corresponding spherical coordinate system ($r, \theta, \varphi$):

$$\psi_{inc} = e^{ik_0 z} = \sum_{l=0}^{+\infty} i^l \sqrt{4\pi(2l+1)} j_l(k_0 r) Y_l^0(\theta) \tag{S.2}$$

where $j_l$ is the $l$-th spherical Bessel function of the first kind, $Y_l^m$ are spherical harmonics, and $k_0 = \sqrt{2m_0 E}/\hbar$ is the wave number. Because of the symmetry of the problem, the Hamiltonian $H$ and the angular momentum operators $L^2$ and $L_z$ form a complete set of commuting observables. The solution of (S.1) can therefore be sought as a linear combination of their common eigenvectors $R_l(r) Y_l^m(\theta, \varphi)$, where $R_l(r)$ is a solution of the radial equation, i.e., a linear combination of spherical Bessel functions $j_l$ and $y_l$ or, equivalently, spherical Hankel function $h_l^{(1)}$ and $h_l^{(2)}$. Setting $m=0$ because of rotational invariance around $\vec{z}$ and taking into account the incident wave and the layered geometry of the structure, we retain the following form for the solution of (S.1):

$$\psi = \sum_{l=0}^{+\infty} i^l \sqrt{4\pi(2l+1)} \rho_l(k_0 r) Y_l^0(\theta) \tag{S.3}$$

with

$$\rho_l(k_0 r) = \begin{cases} j_l(k_0 r) + a_l h_l^{(1)}(k_0 r), & a_{c2} \leq r < +\infty \\ b_l j_l(k_c r) + c_l y_l(k_c r), & a_{c1} \leq r \leq a_{c2} \\ d_l j_l(k_{in} r) + e_l y_l(k_{in} r), & a \leq r \leq a_{c1} \end{cases} \quad (S.4)$$

where we have introduced the scattering coefficients $a_l, b_l, c_l, d_l$, and $e_l$ and the core-shell wave numbers $k_c = \sqrt{2m_0(E-V_c)}/\hbar$, $k_{in} = \sqrt{2m_0(E-V_{in})}/\hbar$. The presence of an *s*-resonator centered at the origin affects only the scattering coefficient $a_0$, while all the $l \geq 1$ partial waves are identical to the case of an empty core-shell structure. The expressions for the $l \geq 1$ coefficients are therefore readily available in previous contributions on quantum cloaking, for instance in [1]. For the $l = 0$ case, we first enforce the boundary conditions at the interfaces $r = a_{c1}$ and $r = a_{c2}$:

$$\rho_0 \text{ and } \frac{1}{m_0}\frac{\partial \rho_0}{\partial r} \text{ continuous,} \quad (S.5)$$

yielding four equations. The last equation is obtained writing the logarithmic derivative boundary condition at $r = a$:

$$\left.\frac{a}{m_0}\frac{u'}{u}\right|_{r=a} = const = q_0 \quad (S.6)$$

where by definition $u(r) = r\rho(r)$, and the energy dependency of $q_0$ is associated to a Breit-Wigner *s*-resonator with resonant energy $E_R$, elastic scattering width $\Gamma_{el}$ and absorption width $\Gamma_{abs}$, i.e., in the vicinity of $E_R$ [2]

$$m_0 q_0(E) = (E - E_R)\left.\frac{d(\text{Re } q_0)}{dE}\right|_{E=E_R} + i \text{Im } q_0, \quad (S.7)$$

with

$$\Gamma_{el} = -\frac{2k_0 a}{\left.\dfrac{d(\mathrm{Re}\, q_0)}{dE}\right|_{E=E_R}} \tag{S.8}$$

and

$$\Gamma_{abs} = \frac{2\,\mathrm{Im}\, q_0}{\left.\dfrac{d(\mathrm{Re}\, q_0)}{dE}\right|_{E=E_R}}. \tag{S.9}$$

After some calculations, one obtains the following linear system

$$A \begin{pmatrix} d_0 \\ e_0 \\ b_0 \\ c_0 \\ a_0 \end{pmatrix} = \begin{pmatrix} 0 \\ 0 \\ 0 \\ j_0(k_0 a_c) \\ \dfrac{k_0}{m_0} j_0'(k_0 a_c) \end{pmatrix}, \tag{S.10}$$

with

$$A = \begin{bmatrix} \dfrac{m_0 Q_0}{k_{in} a} j_0(k_{in} a) - j_0'(k_{in} a) & \dfrac{m_0 Q_0}{k_{in} a} y_0(k_{in} a) - y_0'(k_{in} a) & 0 & 0 & 0 \\ -j_0(k_{in} a_{c1}) & -y_0(k_{in} a_{c1}) & j_0(k_c a_{c1}) & y_0(k_c a_{c1}) & 0 \\ -\dfrac{k_{in}}{m_0} j_0'(k_{in} a_{c1}) & -\dfrac{k_{in}}{m_0} y_0'(k_{in} a_{c1}) & \dfrac{k_c}{m_0} j_0'(k_c a_{c1}) & \dfrac{k_c}{m_0} y_0'(k_c a_{c1}) & 0 \\ 0 & 0 & j_0(k_c a_{c2}) & y_0(k_c a_{c2}) & -h_0^{(1)}(k_0 a_{c2}) \\ 0 & 0 & \dfrac{k_c}{m_0} j_0'(k_c a_{c2}) & \dfrac{k_c}{m_0} y_0'(k_c a_{c2}) & -\dfrac{k_0}{m_0} h_0^{(1)'}(k_0 a_{c2}) \end{bmatrix} \tag{S.11}$$

where $Q_0$ satisfies $m_0 Q_0 = m_0 q_0 - 1$. This linear system can be solved to yield all the scattering coefficients for the *s*-wave.

The coefficients $S_l$ are easily calculated using the relation

$$S_l = 2a_l + 1. \tag{S.12}$$

The results presented in the main text may be reproduced using the formulas above for $m_0 = 9.1 \times 10^{-31}$ kg, $E_R = 0.15\,\text{meV}$, $\Gamma_{el} = 10^{-5}\,\text{eV}$, $\Gamma_{abs} = 10^{-6}\,\text{eV}$, $a = 1\,\text{Å}$, $a_{c2} = 10a$ and $a_{c1} = 0.8 a_{c2}$.